\documentclass[%
 aip,
rsi,%
 amsmath,amssymb,
 reprint,%
]{revtex4-1}

\usepackage{graphicx}% Include figure files
\usepackage{dcolumn}% Align table columns on decimal point
\usepackage{bm}% bold math
\usepackage{verbatim}
\usepackage{mathtools}
\newcounter{tempEquationCounter} 
\newcounter{thisEquationNumber}

\begin{document}

\preprint{AIP/123-QED}

\title[Response of QD to structured beams via convolution integrals]
{Response of QD to structured beams via convolution integrals}% Force line breaks with \\
%\thanks{Footnote to title of article.}

\author{J. Narag}
\email{jnarag@nip.upd.edu.ph}
% \altaffiliation[Also at ]{National Institute of Physics, University of the Philippines, Dilaman, Quezon City, Philippines.}%Lines break automatically or can be forced with \\
\author{N. Hermosa}%
 %\email{Second.Author@institution.edu.}
%\\This line break forced with \textbackslash\textbackslash

% \homepage{http://www.Second.institution.edu/~Charlie.Author.}
%\affiliation{%
%Second institution and/or address%\\This line break forced% with \\
%}%

\date{\today}% It is always \today, today,

\begin{abstract}
We propose a new expression for the response of a quadrant detector using convolution integrals. This expression is easier to evaluate by hand, exploiting the properties of the convolution. Computationally, it is also practicable to use since a large number of computer programs can right away evaluate convolutions. We use the new expression to obtain an analytical form of the response of a quadrant detector to a Gaussian beam and to Hermite-Gaussian beams in general. We compare this analytic expression for the response for the Gaussian beam with the approximations from previous studies and with a response obtained through simulations. From the response, we also obtained an analytical form for the sensitivity of the quadrant detector to a Gaussian beam. Lastly, we demonstrate the computational ease of using our new expression for the response calculating the sensitivity of the quadrant detector to the Bessel beam.
\end{abstract}

%Use showkeys class option if keyword
\maketitle

\section{\label{sec:level1} INTRODUCTION}
Position sensitive devices, such as quadrant detectors, have attracted the interest of the scientific community ever since Putman, in 1992, showed that beam deflection techniques have comparable sensitivities as interferometric methods \cite{putman1992detailed}. To date, quadrants detectors are widely used in atomic force microscopy, image scanning microscopy, laser alignment, space communication, and in optical tweezing, to name a few \cite{putman1992detailed,castello2015image,kuang2004analyzing,zhao2010application,keen2007comparison}. Nonetheless, experimental and theoretical ventures are still being pursued to improve the position estimate and the sensitivity of the quadrant detectors \cite{lu2014novel,wu2015improved, chen2013investigation,cui2010linearity,cui2010improved,olyaee20123}. This is mainly accomplished either by configuring the quadrant detector or by changing the structure of the beam incident on the quadrant detector \cite{barbaric2013sensitivity, esper2016configurable, hao2012quadrant, lee2010detection, salles2010designing, esper2016configurable,nugrowati2012position}. Ironically, the biggest challenges in the advances of quadrant detection are the calculation of the response and the sensitivity of the quadrant detector \cite{manojlovic2011quadrant,hermosa2011quadrant}. This difficulty is principally due to complicated integrals involved in calculating such quantities \cite{lu2014novel,zheng2014theoretical,wu2015improved}. Thus, the analysis of the response and the sensitivity of quadrant detectors have mainly relied on numerical simulations and on approximations such as, if the beam size or the beam displacement is very small compared to the detector size \cite{zheng2014theoretical,hermosa2011quadrant,cui2011analysis}

In this work, we introduce an expression for the response of the quadrant detector by using the convolution. We use this method to obtain analytical expressions for the response of the quadrant detector to the Gaussian beam and Hermite-Gaussian beams. From the response, we also obtain an analytical expression for the the sensitivity for the Gaussian beam. The response for the Gaussian beam is compared with previous approximations in Ref. \cite{hermosa2011quadrant} and \cite{manojlovic2011quadrant}, and with numerical simulations. We also demonstrate the computationally ease of our expression by calculating the response for the Bessel beam.

\section{RESPONSE AS A CONVOLUTION}
The response of the quadrant detector is the position estimate of the laser beam relative to the center of the quadrant detector. It is obtained by comparing the intensities impinging on the 4 quadrants of the detector and is given by,
\begin{equation}
R_x(x,y) = \frac{\left(I_1 + I_4 \right) - \left( I_2 + I_3 \right)}{I_1+I_2+I_3+I_4},
\label{eq:Rx}
\end{equation}
\begin{equation}
R_y(x,y) = \frac{\left(I_1 + I_2 \right) - \left( I_3 + I_4 \right)}{I_1+I_2+I_3+I_4},
\label{eq:Ry}
\end{equation}
\noindent where $R_x(x,y)$ and $R_y(x,y)$ are the responses in the horizontal and vertical directions, respectively, and $I_1(x,y)$ to $I_4(x,y)$ are the intensities on quadrants $1$ to $4$ of the detector, respectively. From \eqref{eq:Rx}, $R_x$ can be understood as the the difference in the intensity between the left and right halves of the detector, scaled by the total intensity on the four quadrants of the detector. Similarly, $R_y$ is the difference between the intensity in the upper and lower halves of the detector, scaled by the total intensity. Although these formulas for the response are simple, they are arduous to evaluate because of the $I_i(x,y)$'s involve integrals over the \textit{ith} quadrant which depend on the beam position and the beam profile. This makes the setting up of the integral equally difficult as evaluating it. Our new expression simplifies these integrals by recasting it in terms of the convolution.

In 1D, the convolution, $h(x)$, of two functions, $f(x)$ and $g(x)$, denoted by,
\begin{equation}
h(x) = f(x) \ast g(x) = \int_{-\infty}^{+\infty} f(x')g(x-x')dx',
\label{eq:1Dconv}
\end{equation}
can be interpreted as the multiplicative overlap of the functions $f(x)$ and $g(x)$ when they are translated $x$ distance away from one another. In 2D, the convolution is given by,
\begin{align}
H(x,y) &= F(x,y) \ast\ast G(x,y) \\
&= \int_{-\infty}^{+\infty}\int_{-\infty}^{+\infty} F(x',y')G(x-x',y-y')dx'dy'.
\end{align}
Again, this can be interpreted as a multiplicative overlap between the functions $F(x,y)$ and $G(x,y)$ when they are separated by a distance $x$ and $y$ in the horizontal and vertical directions, respectively. We can therefore express the intensities on each half on the detector as the convolution of the beam profile and the function describing the corresponding half of the detector,
\begin{align}
\begin{split}
I_1 + I_2 = D_r(x,y)**B(x,y),
\end{split} \\
\begin{split}
I_3 + I_4 = D_l(x,y)**B(x,y),
\end{split}
\end{align}
where $B(x,y)$ is the beam intensity profile and $D_r$ and $D_l$ are the functions corresponding to the right and left halves of the quadrant detector, respectively. For a square detector of unit side length, $D_r$ and $D_l$ are respectively defined as,

\begin{align}
D_r(x,y) &= \begin{cases}
1, & 0\leq x\leq 1,\ \ -1\leq y\leq 1\\
0, & \text{otherwise,}
\label{eq:Dr}
\end{cases}\\
D_l(x,y) &= \begin{cases}
1, & -1\leq x\leq 0,\ \ -1\leq y\leq 1\\
0, & \text{otherwise}.
\label{eq:Dl}
\end{cases}
\end{align}
Using these definitions, Eq. \eqref{eq:Rx} can rewritten as,
\begin{equation}
R_x(x,y) = \frac{D_r(x,y)**B_i(x,y) - D_l ** B_i(x,y)}{D_r(x,y)**B_i(x,y) + D_l ** B_i(x,y)}.
\label{eq:Rcx}
\end{equation}
\noindent Similarly we can rewrite \eqref{eq:Ry} as,
\begin{equation}
R_y(x,y) = \frac{D_t(x,y)**B_i(x,y) - D_b ** B_i(x,y)}{D_t(x,y)**B_i(x,y) + D_b ** B_i(x,y)},
\label{eq:Rcy}
\end{equation}
\noindent where $D_t$ and $D_b$ correspond to the top and the bottom halves of the detector and are, for a square detector, defined as,
\begin{align}
D_t(x,y) &= \begin{cases}
1, & 0\leq y\leq 1,\ \ -1\leq x\leq 1\\
0, & \text{otherwise,}
\label{eq:Dt}
\end{cases}\\
D_b(x,y) &= \begin{cases}
1, & -1\leq y\leq 0,\ \ -1\leq x\leq 1\\
0, & \text{otherwise}.
\label{eq:Db}
\end{cases}
\end{align}

These new equations, Eqs. \eqref{eq:Rcx} and \eqref{eq:Rcy}, for the response looks more intimidating than the original equations, Eqs. \eqref{eq:Rx} and \eqref{eq:Ry}, and it seems like we haven't gained the simplicity we desired. But, by writing the response as convolutions, we have completely eliminated the process of setting up the integral which can be as hard as evaluating the integral themselves. Furthermore, for special cases, the properties of the convolution can be exploited to simplify the evaluation of the integral. In the next sections, we use Eqs. \eqref{eq:Rcx} and \eqref{eq:Rcy} to obtain analytical expressions for the responses for the Gaussian beam and the Hermite-Gaussian beams.

\section{Response and Sensitivity for Gaussian}
Using Eq. \eqref{eq:Rcx}, we calculate the response $R_x$ of the quadrant detector to a Gaussian beam given by,
\begin{equation}
G(x,y) = G_0  \exp \left( -\frac{2}{w_0^2}(x^2+y^2)\right)
\label{eq:G}
\end{equation}
\noindent where $w_0$ is the beam waist and $G_0$ is a normalization constant. Evaluating \eqref{eq:Rcx} for the Gaussian beam gives,
\begin{align}
R_x^G(x,y) &= 
\frac{D_r(x,y) ** G(x,y) - D_l**G(x,y)}{D_r(x,y) ** G(x,y) + D_l**G(x,y)} \\
\begin{split}
&= \Bigg\lbrace \int_{-\infty}^{+\infty} \int_{-\infty}^{+\infty} D_r(x,y) G(x'-x, y'-y)dxdy \\
& \quad\quad- \int_{-\infty}^{+\infty} \int_{-\infty}^{+\infty} D_l(x,y) G(x'-x, y'-y)dxdy \Bigg\rbrace \\ 
&\quad\div \Bigg\lbrace \int_{-\infty}^{+\infty} \int_{-\infty}^{+\infty} D_r(x,y) G(x'-x, y'-y)dxdy \\
&\quad\quad + \int_{-\infty}^{+\infty} \int_{-\infty}^{+\infty} D_l(x,y) G(x'-x, y'-y)dxdy \Bigg\rbrace
\label{eq:RxIG}
\end{split}\\
\begin{split}
R_x^G &= \frac{2\text{erf}\left(\sqrt{2}\frac{x}{w_0}\right)-\text{erf}\left(\sqrt{2}\frac{x+1}{w_0}\right)-\text{erf}\left(\sqrt{2}\frac{x-1}{w_0}\right)}{\text{erf}\left(\sqrt{2}\frac{x+1}{w_0}\right)-\text{erf}\left(\sqrt{2}\frac{x-1}{w_0}\right)}
\end{split}
\label{eq:Rxg}
\end{align}
where $\text{erf}(x)$ is the error function. This demonstrates how much easier it is to calculate the the response as we did not have to deal with setting up the $I_i(x,y)$'s. Additionally, the infinite integrals in the convolution are reduced to finite integrals since the detector functions, $D_r$ and $D_l$ are unity and are non-zero only over the finite regions $0 \leq x \leq 1, -1 \leq y\leq 1$ and $-1 \leq x \leq 0, -1 \leq y\leq 1$, respectively. Moreover, we did not have to evaluate the y-integrals because the symmetry of the Gaussian beam and the detector functions allowed the separation of the 2D integrals into two separate integrals in x and y, and the y-integrals cancel out. Thus the response, $R_x^G$, only depends on $x$ as expected. This result that the response should not depend on $y$ agrees with intuition since along any horizontal line the intensity distribution remains a Gaussian. Also because of the symmetry, the response in the y-direction, $R_y^G(x,y)$, is the same as \eqref{eq:Rxg} but with x replaced with y and with y replaced with x,
\begin{align}
\begin{split}
R_y^G &= \frac{2 \text{erf}\left(\sqrt{2}\frac{y}{w_0}\right)-\text{erf}\left(\sqrt{2}\frac{y+1}{w_0}\right)-\text{erf}\left(\sqrt{2}\frac{y-1}{w_0}\right)}{\text{erf}\left(\sqrt{2}\frac{y+1}{w_0}\right)-\text{erf}\left(\sqrt{2}\frac{y-1}{w_0}\right)}.
\end{split}
\end{align}

In the calculations above, we have implicitly defined the beam displacement relative to the detector size when we defined the detector functions in Eqs. \eqref{eq:Dr} to \eqref{eq:Dl} and  \eqref{eq:Dt} to \eqref{eq:Db}. That is, the detector side length is 2 units.

The sensitivity is then obtained by differentiating the response, $R^G_X$ with respect to the $x$,
\begin{align}
S_x &= \frac{dR_x^G}{dx}\\
\begin{split}
&= \frac{4}{w_o}\sqrt{\frac{2}{\pi}} \Bigg\lbrace exp\left( \frac{\sqrt{2}(x)}{w_o}\right)\text{erf}\left(\frac{-2(x^2)}{w_o^2}\right)\\ 
&\quad\quad\quad+ exp\left( \frac{\sqrt{2}(x+1)}{w_o}\right)\text{erf}\left(\frac{-2((x-1)^2)}{w_o^2}\right)\\ 
&\quad\quad\quad- exp\left( \frac{\sqrt{2}(x-1)}{w_o}\right)\text{erf}\left(\frac{-2((x+1)^2)}{w_o^2}\right) \Bigg\rbrace  \div \\
&\quad\quad\quad{\Bigg\lbrace \text{erf}\left(\frac{\sqrt{2}(x+1)}{w_o} \right) - \text{erf}\left(\frac{\sqrt{2}(x-1)}{w_o} \right)\Bigg\rbrace }^2.\\
\end{split}
\end{align}

We have only calculated the sensitivities in the x-directions. However, due to the symmetry of the Gaussian beam $S_y(y) = S_x(x)$. Also, notice that the sensitivity in $x$ is independent of $y$. This means that the alignment of the beam along the y-axis will not have any effect on the sensitivity along the x-axis, and vice versa. This is true only for beams that have separable $x$ and $y$ integrals as we will show later.

Additionally, we can also compute for the response of a  general Hermite-Gaussian beam, with intensity given by,
\begin{equation}
I_{lm}= I_o H_l^2\left(\frac{\sqrt{2}x}{w}\right) H_m^2\left(\frac{\sqrt{2}y}{w}\right) \exp\left(\frac{-2(x+y)^2}{w^2}\right),
\end{equation}
where $I_o$ is a normalization, $\omega$ is the beam waist and $H_n(x)$ are the Hermite polynomials, \eqref{eq:Rcx} gives,

\begin{widetext}
\begin{equation}
R_{hg_{l=0}} = 
\left[\dfrac{\splitdfrac{\sum_{v=0}^l \sqrt{\frac{\pi}{2}} 2^{2l-v} v! {\binom{l}{v}}^2\Bigg\lbrace \frac{2x _1F_1(l-v+\frac{1}{1};\frac{3}{2};\frac{-2x^2}{w^2}) - (x+1) _1F_1(l-v+\frac{1}{1};\frac{3}{2};\frac{-2(x+1)^2}{w^2}) - (x-1) _1F_1(l-v+\frac{1}{1};\frac{3}{2};\frac{-2(x-1)^2}{w^2})}{\frac{w}{\sqrt2}\Gamma\left(\frac{1}{2}+v-l\right)} }{+ \frac{2 _1F_1(l-v;\frac{1}{2}; \frac{-2x^2}{w^2}) - _1F_1(l-v;\frac{1}{2}; \frac{-2(x+1)^2}{w^2}) - _1F_1(l-v;\frac{1}{2}; \frac{-2(x-1)^2}{w^2}) }{2(l-v)\Gamma(v-l)} \Bigg\rbrace}}
{\splitdfrac{\sum_{v=0}^l \sqrt{\frac{\pi}{2}} 2^{2l-v} v! {\binom{l}{v}}^2 \Bigg\lbrace \frac{(x+1) _1F_1(l-v+\frac{1}{2};\frac{3}{2}; \frac{-2(x+1)^2}{w^2}) -   (x+1) _1F_1(l-v+\frac{1}{2};\frac{3}{2}; \frac{-2(x+1)^2}{w^2}) }{\frac{w}{\sqrt2} \Gamma\left(\frac{1}{2}+v-l\right)}}{+ \frac{ _1F_1(l-v;\frac{1}{2};\frac{-2(x+1)^2}{w^2}) - _1F_1(l-v;\frac{1}{2};\frac{-2(x-1)^2}{w^2}) } {2(l-v)\Gamma\left(v-l\right)} \Bigg\rbrace}}\right]
\label{eq:Rhg}
\end{equation}
\end{widetext}

Again $R_{x}^{HG}$ is independent not only on $y$, but also on the the mode, $m$, of the Hermite polynomial that envelops the $y$ direction. This is a less intuitive result, since the intensity along horizontal lines across HG beams with arbitrary modes is not constant. The resolution to this is that, just like for the Gaussian beam, the 2D convolution for the HG beam is also separable into two integrals in $x$ and in $y$ and the division kills out the $y$ integrals. For $l=0$, the summations in $R_{x}^{HG}$ reduce to single terms and with the following identities,
\begin{equation}
erf(x) = \frac{2x}{\sqrt\pi} _1F_1(\frac{1}{2},\frac{3}{2},-x^2),
\end{equation}
\begin{equation}
_1F_1(0,\frac{1}{2}, -x^2) = 1,
\end{equation}
\begin{equation}
\lim_{x\to0} x\Gamma(x) = 1,
\end{equation}
we obtain the the response for the Gaussian beam in  Eq. \eqref{eq:Rxg}. In the next section, we compare the response of the Gaussian beam to previous studies and to simulations and also computationally obtain the response for a Bessel beam.

\section{Numerical Results}
We simulated the response of a quadrant detector to a Gaussian beam by generating an $n$ x $n$ image of the Gaussian beam, then taking the difference between the sum of the pixel values in the left and in the right half of the image and dividing by the total pixel value of the whole image,
\begin{equation}
R_x(x) = \frac{\sum\limits_{j=-n}^n \sum\limits_{i=\frac{n}{2}}^n I_{ji} - \sum\limits_{j=-n}^n \sum\limits_{i=1}^{n} I_{ji} }{\sum\limits_{j=-n}^n \sum\limits_{i=\frac{n}{2}}^n I_{ji} + \sum\limits_{j=-n}^n \sum\limits_{i=1}^{n} I_{ji} }
\label{eq:R_dig}
\end{equation}
where $I_{ji}$ is the pixel value of the j-ith element of the image. The Gaussian beam was then translated along the x-axis to obtain the response for different beam displacements. Figure 1 illustrates the simulation for a Bessel beam. 
\begin{figure}
\includegraphics[width=\linewidth]{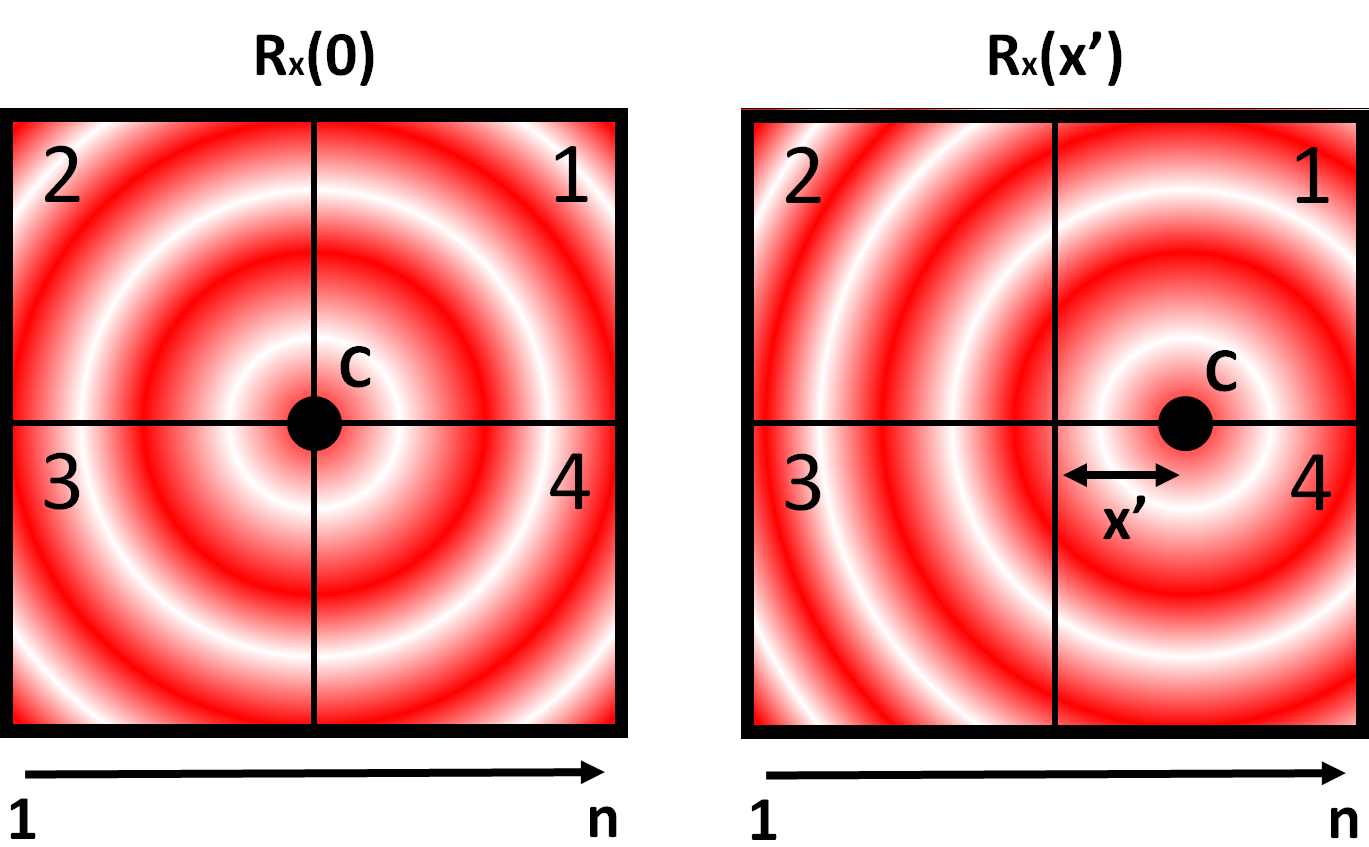}% Here is how to import EPS art
\caption{Simulation process of obtaining the response by applying Eq. \eqref{eq:Rx} thru Eq. \eqref{eq:R_dig}. The intensity differences and sums of the quadrants are obtained to determine the response and sensitivity with displacement of the QD.}
\end{figure}

\begin{figure}[b]
\centering
\includegraphics[width=\linewidth]{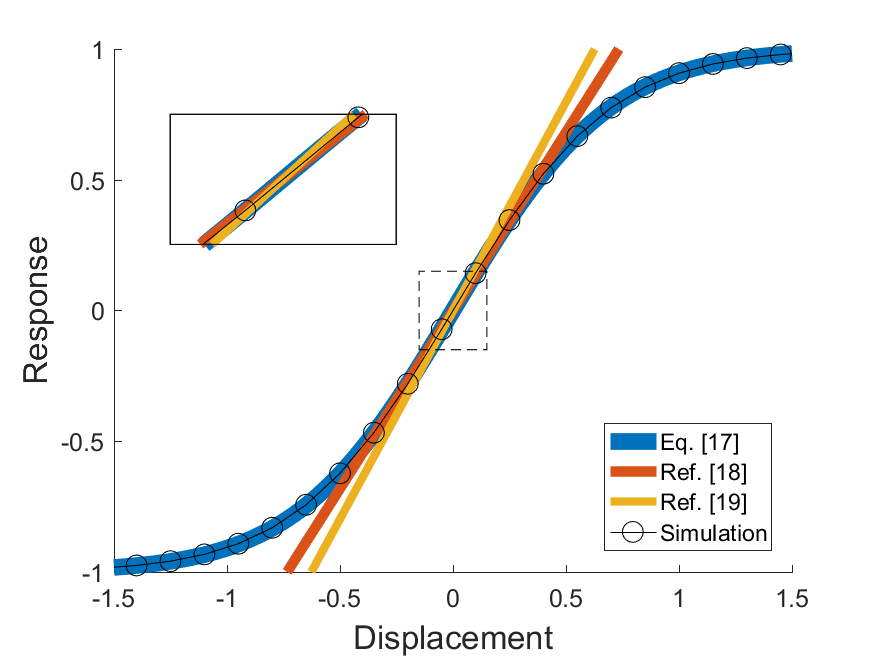}
\caption{Response $R_x$ as a function of x-displacement relative to the detector's center. Our derived expression for $R_x$ agrees well with the simulated response from Eq. \eqref{eq:R_dig} and in approximations due to \cite{hermosa2011quadrant,manojlovic2011quadrant}.}
\label{fig:response}
\end{figure}

This process of simulating the response arises from the definition given by Eq. \eqref{eq:Rx}. We compared this with the analytical expression Eq. \eqref{eq:Rcx} by plotting them in Fig. \ref{fig:response}. Also included in Fig. \ref{fig:response} are the approximations for the response from the results of Ref. [18] and Ref. [19]. We used the result with $l=0$ to correspond to a Gaussian beam in Ref. [19]. Reference [18] derived an approximate form, not for the response, but for the sensitivity for the Gaussian beam. We obtained the corresponding response by straightforward integration. As shown in Fig. \ref{fig:response}, our expression for the response in Eq. \eqref{eq:Rxg}, agrees well with the numerical results. While the approximations from Ref. [18] and Ref. [19] are only acceptable for small displacements as shown in the inset. 

\begin{figure}[t]
\centering
\includegraphics[width=\linewidth]{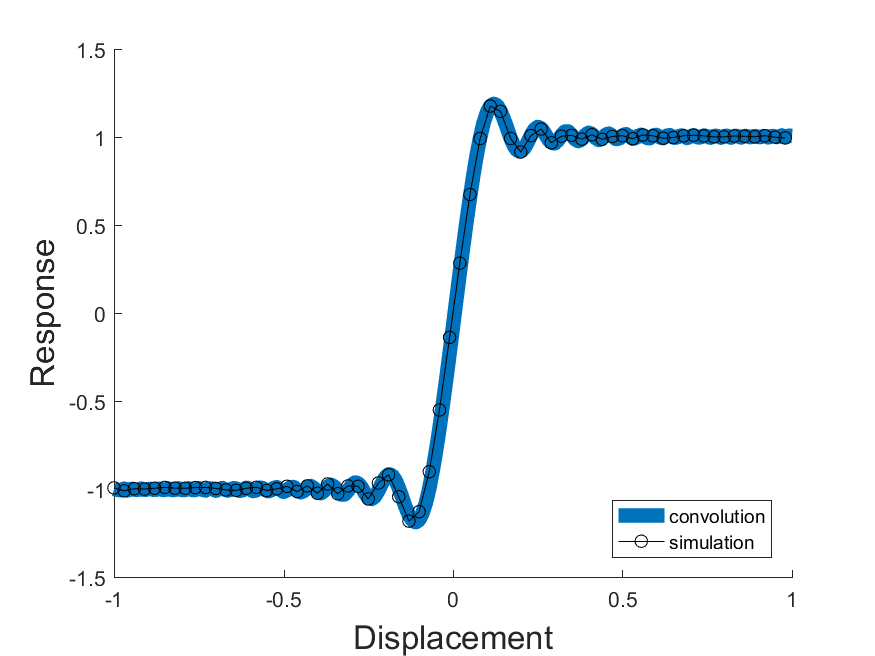}
\caption{Response $R_x$ as a function of x- displacement for the Bessel beam when the beam is centered in y. Our expression from the convolution integrals gives similar result with the simulation.}
\label{fig:Bessel}
\end{figure}

To illustrate the computational ease of using Eq. \eqref{eq:Rcx}, we apply it to a Bessel beam given by, 
\begin{equation} 
J(r,\theta) = B_0 \:\:{J_0}^2(kr),
\end{equation}
where $J_0(r)$ is the zeroth-order Bessel function, $k$ is the wave vector and $B_0$ is a normalization. Since the Bessel beam is more naturally defined in polar coordinates, it will be hard to compute its response by hand. However, we can let the computer evaluate Eq. \eqref{eq:Rcx} with a few lines of code. Figure \ref{fig:Bessel} shows the response in $x$ of the quadrant detector to the Bessel beam obtained by computationally applying Eq. \eqref{eq:Rcx}, shown in orange, and by simulation using Eq. \eqref{eq:Rx}, shown in blue. The simulation for obtaining the response was the same as described previously for the Gaussian beam. Again, it is obvious from the plot, that they are equivalent to each other. The main difference is that Eq. \eqref{eq:Rcx} is more straightforward to apply computationally than to do the simulation based on Eq. \eqref{eq:Rx}. Another advantage to using Eq. \eqref{eq:Rcx} over Eq. \eqref{eq:Rx}, is that it automatically gives the response as a function of $x$ and $y$ as shown in Fig. \ref{fig:Bessel}. Note that, in contrast to the Gaussian and HG beam, the response for the Bessel depends in both $x$ and $y$, as depicted by the non-symmetry in $y$. That is, there is a slight curving in the lines of constant response instead of straight vertical lines for a Gaussian beam. This means the the vertical position response depends on the horizontal alignment of the beam. The reason is because the intensity distribution of the Bessel cannot be written as a product $f(x)g(y)$, thus the 2D convolution cannot be decomposed into two integrals in x and y and the response is a function of both x and y.

\begin{figure}[h]
\centering
\includegraphics[width=\linewidth]{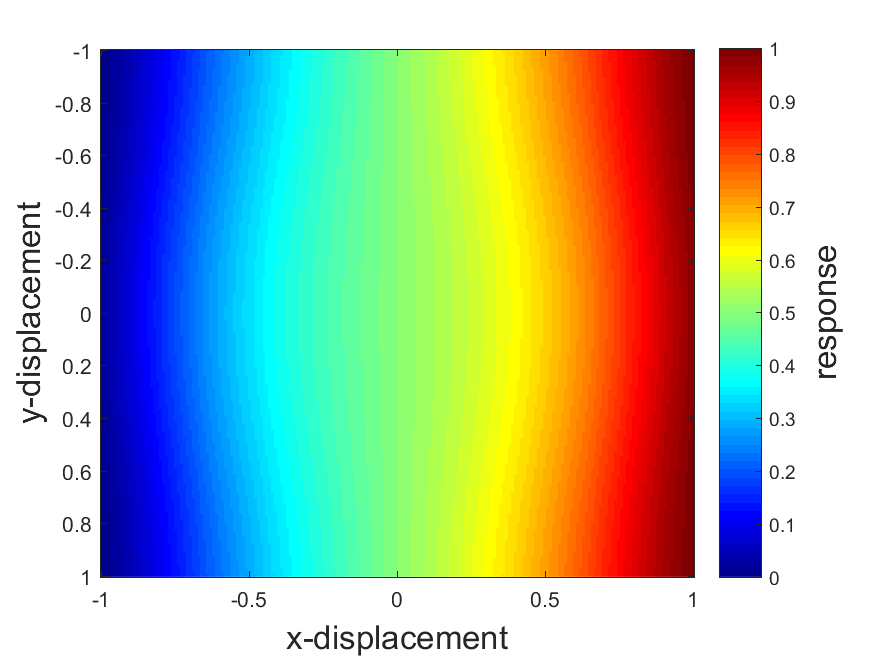}
\caption{Response $R_x$ as a function of x and y-displacement for the Bessel beam. When the beam is offset in the y-direction, the $R_x$ changes and hence the $S_x$ also differs. }
\label{fig:Bessel01}
\end{figure}

\section{Conclusion}
We present an expression for the response of the quadrant detector using convolution integrals in this paper. This expression is easier to evaluate by hand because it skips the process of setting up the integral which can be as hard the the evaluating it. Computationally, it is also more straightforward to execute since the convolution is directly calculated by many different softwares. We use our expression to obtain analytical forms of the response for the Gaussian beam and HG beam.  We also demonstrated the computational ease for our method by calculating the response of the Bessel beam. We show that for beams that have separable $x$ and $y$ integrals, the response and hence, the sensitivity for a certain axis is independent of the beam's position perpendicular to that axis where it is being displaced. Since Bessel beams have non-separable $x$ and $y$ integrals, the response and sensitivity changes with the beam's position. Our results are important in position measurements using quadrant detectors and in the optimization of the quadrant detector for various applications.

\begin{acknowledgments}
The authors acknowledge the support of the University of the Philippines Office of the Vice President for Academic Affairs thru the Balik PhD Program (OVPAA Balik PhD 2015-06) and the Enhanced Creative Work and Research Grant (ECWRG 2016-2-027) and the Philippine Council for Industry, Energy and Emerging Technology Research and Developmen (PCIEERD) of Department of Science and Technology of the Republic of the Philippines.
\end{acknowledgments}

\nocite{*}
\bibliography{aipsamp}% Produces the bibliography via BibTeX.

\begin{thebibliography}{21}%
\makeatletter
\providecommand \@ifxundefined [1]{%
 \@ifx{#1\undefined}
}%
\providecommand \@ifnum [1]{%
 \ifnum #1\expandafter \@firstoftwo
 \else \expandafter \@secondoftwo
 \fi
}%
\providecommand \@ifx [1]{%
 \ifx #1\expandafter \@firstoftwo
 \else \expandafter \@secondoftwo
 \fi
}%
\providecommand \natexlab [1]{#1}%
\providecommand \enquote  [1]{``#1''}%
\providecommand \bibnamefont  [1]{#1}%
\providecommand \bibfnamefont [1]{#1}%
\providecommand \citenamefont [1]{#1}%
\providecommand \href@noop [0]{\@secondoftwo}%
\providecommand \href [0]{\begingroup \@sanitize@url \@href}%
\providecommand \@href[1]{\@@startlink{#1}\@@href}%
\providecommand \@@href[1]{\endgroup#1\@@endlink}%
\providecommand \@sanitize@url [0]{\catcode `\\12\catcode `\$12\catcode
  `\&12\catcode `\#12\catcode `\^12\catcode `\_12\catcode `\%12\relax}%
\providecommand \@@startlink[1]{}%
\providecommand \@@endlink[0]{}%
\providecommand \url  [0]{\begingroup\@sanitize@url \@url }%
\providecommand \@url [1]{\endgroup\@href {#1}{\urlprefix }}%
\providecommand \urlprefix  [0]{URL }%
\providecommand \Eprint [0]{\href }%
\providecommand \doibase [0]{http://dx.doi.org/}%
\providecommand \selectlanguage [0]{\@gobble}%
\providecommand \bibinfo  [0]{\@secondoftwo}%
\providecommand \bibfield  [0]{\@secondoftwo}%
\providecommand \translation [1]{[#1]}%
\providecommand \BibitemOpen [0]{}%
\providecommand \bibitemStop [0]{}%
\providecommand \bibitemNoStop [0]{.\EOS\space}%
\providecommand \EOS [0]{\spacefactor3000\relax}%
\providecommand \BibitemShut  [1]{\csname bibitem#1\endcsname}%
\let\auto@bib@innerbib\@empty
%</preamble>
\bibitem [{\citenamefont {Putman}\ \emph {et~al.}(1992)\citenamefont {Putman},
  \citenamefont {De~Grooth}, \citenamefont {Van~Hulst},\ and\ \citenamefont
  {Greve}}]{putman1992detailed}%
  \BibitemOpen
  \bibfield  {author} {\bibinfo {author} {\bibfnamefont {C.~A.}\ \bibnamefont
  {Putman}}, \bibinfo {author} {\bibfnamefont {B.~G.}\ \bibnamefont
  {De~Grooth}}, \bibinfo {author} {\bibfnamefont {N.~F.}\ \bibnamefont
  {Van~Hulst}}, \ and\ \bibinfo {author} {\bibfnamefont {J.}~\bibnamefont
  {Greve}},\ }\href@noop {} {\bibfield  {journal} {\bibinfo  {journal} {Journal
  of Applied Physics}\ }\textbf {\bibinfo {volume} {72}},\ \bibinfo {pages} {6}
  (\bibinfo {year} {1992})}\BibitemShut {NoStop}%
\bibitem [{\citenamefont {Castello}\ \emph {et~al.}(2015)\citenamefont
  {Castello}, \citenamefont {Sheppard}, \citenamefont {Diaspro},\ and\
  \citenamefont {Vicidomini}}]{castello2015image}%
  \BibitemOpen
  \bibfield  {author} {\bibinfo {author} {\bibfnamefont {M.}~\bibnamefont
  {Castello}}, \bibinfo {author} {\bibfnamefont {C.~J.}\ \bibnamefont
  {Sheppard}}, \bibinfo {author} {\bibfnamefont {A.}~\bibnamefont {Diaspro}}, \
  and\ \bibinfo {author} {\bibfnamefont {G.}~\bibnamefont {Vicidomini}},\
  }\href@noop {} {\bibfield  {journal} {\bibinfo  {journal} {Optics letters}\
  }\textbf {\bibinfo {volume} {40}},\ \bibinfo {pages} {5355} (\bibinfo {year}
  {2015})}\BibitemShut {NoStop}%
\bibitem [{\citenamefont {KUANG}\ \emph {et~al.}(2004)\citenamefont {KUANG},
  \citenamefont {FENG}, \citenamefont {FENG},\ and\ \citenamefont
  {LIU}}]{kuang2004analyzing}%
  \BibitemOpen
  \bibfield  {author} {\bibinfo {author} {\bibfnamefont {C.-f.}\ \bibnamefont
  {KUANG}}, \bibinfo {author} {\bibfnamefont {Q.-b.}\ \bibnamefont {FENG}},
  \bibinfo {author} {\bibfnamefont {J.-y.}\ \bibnamefont {FENG}}, \ and\
  \bibinfo {author} {\bibfnamefont {B.}~\bibnamefont {LIU}},\ }\href@noop {}
  {\bibfield  {journal} {\bibinfo  {journal} {Optical Technique}\ }\textbf
  {\bibinfo {volume} {4}},\ \bibinfo {pages} {001} (\bibinfo {year}
  {2004})}\BibitemShut {NoStop}%
\bibitem [{\citenamefont {ZHAO}\ \emph {et~al.}(2010)\citenamefont {ZHAO},
  \citenamefont {TONG}, \citenamefont {LIU},\ and\ \citenamefont
  {JIANG}}]{zhao2010application}%
  \BibitemOpen
  \bibfield  {author} {\bibinfo {author} {\bibfnamefont {X.}~\bibnamefont
  {ZHAO}}, \bibinfo {author} {\bibfnamefont {S.-f.}\ \bibnamefont {TONG}},
  \bibinfo {author} {\bibfnamefont {Y.-q.}\ \bibnamefont {LIU}}, \ and\
  \bibinfo {author} {\bibfnamefont {H.-l.}\ \bibnamefont {JIANG}},\ }\href@noop
  {} {\bibfield  {journal} {\bibinfo  {journal} {Journal of Optoelectronics.
  Laser}\ }\textbf {\bibinfo {volume} {1}},\ \bibinfo {pages} {015} (\bibinfo
  {year} {2010})}\BibitemShut {NoStop}%
\bibitem [{\citenamefont {Keen}\ \emph {et~al.}(2007)\citenamefont {Keen},
  \citenamefont {Leach}, \citenamefont {Gibson},\ and\ \citenamefont
  {Padgett}}]{keen2007comparison}%
  \BibitemOpen
  \bibfield  {author} {\bibinfo {author} {\bibfnamefont {S.}~\bibnamefont
  {Keen}}, \bibinfo {author} {\bibfnamefont {J.}~\bibnamefont {Leach}},
  \bibinfo {author} {\bibfnamefont {G.}~\bibnamefont {Gibson}}, \ and\ \bibinfo
  {author} {\bibfnamefont {M.}~\bibnamefont {Padgett}},\ }\href@noop {}
  {\bibfield  {journal} {\bibinfo  {journal} {Journal of Optics A: Pure and
  Applied Optics}\ }\textbf {\bibinfo {volume} {9}},\ \bibinfo {pages} {S264}
  (\bibinfo {year} {2007})}\BibitemShut {NoStop}%
\bibitem [{\citenamefont {Lu}\ \emph {et~al.}(2014)\citenamefont {Lu},
  \citenamefont {Zhai}, \citenamefont {Wang}, \citenamefont {Guo},
  \citenamefont {Du},\ and\ \citenamefont {Yang}}]{lu2014novel}%
  \BibitemOpen
  \bibfield  {author} {\bibinfo {author} {\bibfnamefont {C.}~\bibnamefont
  {Lu}}, \bibinfo {author} {\bibfnamefont {Y.-S.}\ \bibnamefont {Zhai}},
  \bibinfo {author} {\bibfnamefont {X.-J.}\ \bibnamefont {Wang}}, \bibinfo
  {author} {\bibfnamefont {Y.-Y.}\ \bibnamefont {Guo}}, \bibinfo {author}
  {\bibfnamefont {Y.-X.}\ \bibnamefont {Du}}, \ and\ \bibinfo {author}
  {\bibfnamefont {G.-S.}\ \bibnamefont {Yang}},\ }\href@noop {} {\bibfield
  {journal} {\bibinfo  {journal} {Optik-International Journal for Light and
  Electron Optics}\ }\textbf {\bibinfo {volume} {125}},\ \bibinfo {pages}
  {3519} (\bibinfo {year} {2014})}\BibitemShut {NoStop}%
\bibitem [{\citenamefont {Wu}\ \emph {et~al.}(2015)\citenamefont {Wu},
  \citenamefont {Chen}, \citenamefont {Gao}, \citenamefont {Li},\ and\
  \citenamefont {Wu}}]{wu2015improved}%
  \BibitemOpen
  \bibfield  {author} {\bibinfo {author} {\bibfnamefont {J.}~\bibnamefont
  {Wu}}, \bibinfo {author} {\bibfnamefont {Y.}~\bibnamefont {Chen}}, \bibinfo
  {author} {\bibfnamefont {S.}~\bibnamefont {Gao}}, \bibinfo {author}
  {\bibfnamefont {Y.}~\bibnamefont {Li}}, \ and\ \bibinfo {author}
  {\bibfnamefont {Z.}~\bibnamefont {Wu}},\ }\href@noop {} {\bibfield  {journal}
  {\bibinfo  {journal} {Applied Optics}\ }\textbf {\bibinfo {volume} {54}},\
  \bibinfo {pages} {8049} (\bibinfo {year} {2015})}\BibitemShut {NoStop}%
\bibitem [{\citenamefont {Chen}\ \emph {et~al.}(2013)\citenamefont {Chen},
  \citenamefont {Yang}, \citenamefont {Jia},\ and\ \citenamefont
  {Gao}}]{chen2013investigation}%
  \BibitemOpen
  \bibfield  {author} {\bibinfo {author} {\bibfnamefont {M.}~\bibnamefont
  {Chen}}, \bibinfo {author} {\bibfnamefont {Y.}~\bibnamefont {Yang}}, \bibinfo
  {author} {\bibfnamefont {X.}~\bibnamefont {Jia}}, \ and\ \bibinfo {author}
  {\bibfnamefont {H.}~\bibnamefont {Gao}},\ }\href@noop {} {\bibfield
  {journal} {\bibinfo  {journal} {Optik-International Journal for Light and
  Electron Optics}\ }\textbf {\bibinfo {volume} {124}},\ \bibinfo {pages}
  {6806} (\bibinfo {year} {2013})}\BibitemShut {NoStop}%
\bibitem [{\citenamefont {Cui}\ and\ \citenamefont
  {Soh}(2010{\natexlab{a}})}]{cui2010linearity}%
  \BibitemOpen
  \bibfield  {author} {\bibinfo {author} {\bibfnamefont {S.}~\bibnamefont
  {Cui}}\ and\ \bibinfo {author} {\bibfnamefont {Y.~C.}\ \bibnamefont {Soh}},\
  }\href@noop {} {\bibfield  {journal} {\bibinfo  {journal} {IEEE Transactions
  on Electron Devices}\ }\textbf {\bibinfo {volume} {57}},\ \bibinfo {pages}
  {2310} (\bibinfo {year} {2010}{\natexlab{a}})}\BibitemShut {NoStop}%
\bibitem [{\citenamefont {Cui}\ and\ \citenamefont
  {Soh}(2010{\natexlab{b}})}]{cui2010improved}%
  \BibitemOpen
  \bibfield  {author} {\bibinfo {author} {\bibfnamefont {S.}~\bibnamefont
  {Cui}}\ and\ \bibinfo {author} {\bibfnamefont {Y.~C.}\ \bibnamefont {Soh}},\
  }\href@noop {} {\bibfield  {journal} {\bibinfo  {journal} {Applied Physics
  Letters}\ }\textbf {\bibinfo {volume} {96}},\ \bibinfo {pages} {081102}
  (\bibinfo {year} {2010}{\natexlab{b}})}\BibitemShut {NoStop}%
\bibitem [{\citenamefont {Olyaee}\ and\ \citenamefont
  {Rezazadeh}(2012)}]{olyaee20123}%
  \BibitemOpen
  \bibfield  {author} {\bibinfo {author} {\bibfnamefont {S.}~\bibnamefont
  {Olyaee}}\ and\ \bibinfo {author} {\bibfnamefont {M.}~\bibnamefont
  {Rezazadeh}},\ }\href@noop {} {\bibfield  {journal} {\bibinfo  {journal}
  {International Journal of Engineering Research and Applications}\ }\textbf
  {\bibinfo {volume} {2}},\ \bibinfo {pages} {1157} (\bibinfo {year}
  {2012})}\BibitemShut {NoStop}%
\bibitem [{\citenamefont {Barbari{\'c}}(2013)}]{barbaric2013sensitivity}%
  \BibitemOpen
  \bibfield  {author} {\bibinfo {author} {\bibfnamefont {{\v{Z}}.}~\bibnamefont
  {Barbari{\'c}}},\ }\href@noop {} {\bibfield  {journal} {\bibinfo  {journal}
  {Scientific Publications of the State University of Novi Pazar Series A:
  Applied Mathematics, Informatics and mechanics}\ }\textbf {\bibinfo {volume}
  {5}},\ \bibinfo {pages} {85} (\bibinfo {year} {2013})}\BibitemShut {NoStop}%
\bibitem [{\citenamefont {Esper-Cha{\'\i}n}\ \emph {et~al.}(2016)\citenamefont
  {Esper-Cha{\'\i}n}, \citenamefont {Escuela}, \citenamefont {Fari{\~n}a},\
  and\ \citenamefont {Sendra}}]{esper2016configurable}%
  \BibitemOpen
  \bibfield  {author} {\bibinfo {author} {\bibfnamefont {R.}~\bibnamefont
  {Esper-Cha{\'\i}n}}, \bibinfo {author} {\bibfnamefont {A.~M.}\ \bibnamefont
  {Escuela}}, \bibinfo {author} {\bibfnamefont {D.}~\bibnamefont {Fari{\~n}a}},
  \ and\ \bibinfo {author} {\bibfnamefont {J.~R.}\ \bibnamefont {Sendra}},\
  }\href@noop {} {\bibfield  {journal} {\bibinfo  {journal} {IEEE Sensors
  Journal}\ }\textbf {\bibinfo {volume} {16}},\ \bibinfo {pages} {109}
  (\bibinfo {year} {2016})}\BibitemShut {NoStop}%
\bibitem [{\citenamefont {Hao}\ \emph {et~al.}(2012)\citenamefont {Hao},
  \citenamefont {Kuang}, \citenamefont {Ku}, \citenamefont {Liu},\ and\
  \citenamefont {Li}}]{hao2012quadrant}%
  \BibitemOpen
  \bibfield  {author} {\bibinfo {author} {\bibfnamefont {X.}~\bibnamefont
  {Hao}}, \bibinfo {author} {\bibfnamefont {C.}~\bibnamefont {Kuang}}, \bibinfo
  {author} {\bibfnamefont {Y.}~\bibnamefont {Ku}}, \bibinfo {author}
  {\bibfnamefont {X.}~\bibnamefont {Liu}}, \ and\ \bibinfo {author}
  {\bibfnamefont {Y.}~\bibnamefont {Li}},\ }\href@noop {} {\bibfield  {journal}
  {\bibinfo  {journal} {Optik-International Journal for Light and Electron
  Optics}\ }\textbf {\bibinfo {volume} {123}},\ \bibinfo {pages} {2238}
  (\bibinfo {year} {2012})}\BibitemShut {NoStop}%
\bibitem [{\citenamefont {Lee}\ \emph {et~al.}(2010)\citenamefont {Lee},
  \citenamefont {Park}, \citenamefont {Kim},\ and\ \citenamefont
  {Kouh}}]{lee2010detection}%
  \BibitemOpen
  \bibfield  {author} {\bibinfo {author} {\bibfnamefont {E.~J.}\ \bibnamefont
  {Lee}}, \bibinfo {author} {\bibfnamefont {Y.}~\bibnamefont {Park}}, \bibinfo
  {author} {\bibfnamefont {C.~S.}\ \bibnamefont {Kim}}, \ and\ \bibinfo
  {author} {\bibfnamefont {T.}~\bibnamefont {Kouh}},\ }\href@noop {} {\bibfield
   {journal} {\bibinfo  {journal} {Current Applied Physics}\ }\textbf {\bibinfo
  {volume} {10}},\ \bibinfo {pages} {834} (\bibinfo {year} {2010})}\BibitemShut
  {NoStop}%
\bibitem [{\citenamefont {Salles}\ and\ \citenamefont
  {de~Lima~Monteiro}(2010)}]{salles2010designing}%
  \BibitemOpen
  \bibfield  {author} {\bibinfo {author} {\bibfnamefont {L.~P.}\ \bibnamefont
  {Salles}}\ and\ \bibinfo {author} {\bibfnamefont {D.~W.}\ \bibnamefont
  {de~Lima~Monteiro}},\ }\href@noop {} {\bibfield  {journal} {\bibinfo
  {journal} {IEEE sensors Journal}\ }\textbf {\bibinfo {volume} {10}},\
  \bibinfo {pages} {286} (\bibinfo {year} {2010})}\BibitemShut {NoStop}%
\bibitem [{\citenamefont {Nugrowati}, \citenamefont {Stam},\ and\ \citenamefont
  {Woerdman}(2012)}]{nugrowati2012position}%
  \BibitemOpen
  \bibfield  {author} {\bibinfo {author} {\bibfnamefont {A.}~\bibnamefont
  {Nugrowati}}, \bibinfo {author} {\bibfnamefont {W.}~\bibnamefont {Stam}}, \
  and\ \bibinfo {author} {\bibfnamefont {J.}~\bibnamefont {Woerdman}},\
  }\href@noop {} {\bibfield  {journal} {\bibinfo  {journal} {Optics express}\
  }\textbf {\bibinfo {volume} {20}},\ \bibinfo {pages} {27429} (\bibinfo {year}
  {2012})}\BibitemShut {NoStop}%
\bibitem [{\citenamefont {Manojlovi{\'c}}(2011)}]{manojlovic2011quadrant}%
  \BibitemOpen
  \bibfield  {author} {\bibinfo {author} {\bibfnamefont {L.~M.}\ \bibnamefont
  {Manojlovi{\'c}}},\ }\href@noop {} {\bibfield  {journal} {\bibinfo  {journal}
  {Applied optics}\ }\textbf {\bibinfo {volume} {50}},\ \bibinfo {pages} {3461}
  (\bibinfo {year} {2011})}\BibitemShut {NoStop}%
\bibitem [{\citenamefont {Hermosa}, \citenamefont {Aiello},\ and\ \citenamefont
  {Woerdman}(2011)}]{hermosa2011quadrant}%
  \BibitemOpen
  \bibfield  {author} {\bibinfo {author} {\bibfnamefont {N.}~\bibnamefont
  {Hermosa}}, \bibinfo {author} {\bibfnamefont {A.}~\bibnamefont {Aiello}}, \
  and\ \bibinfo {author} {\bibfnamefont {J.}~\bibnamefont {Woerdman}},\
  }\href@noop {} {\bibfield  {journal} {\bibinfo  {journal} {Optics letters}\
  }\textbf {\bibinfo {volume} {36}},\ \bibinfo {pages} {409} (\bibinfo {year}
  {2011})}\BibitemShut {NoStop}%
\bibitem [{\citenamefont {Zheng}, \citenamefont {Li},\ and\ \citenamefont
  {Zhang}(2014)}]{zheng2014theoretical}%
  \BibitemOpen
  \bibfield  {author} {\bibinfo {author} {\bibfnamefont {Z.}~\bibnamefont
  {Zheng}}, \bibinfo {author} {\bibfnamefont {C.}~\bibnamefont {Li}}, \ and\
  \bibinfo {author} {\bibfnamefont {S.}~\bibnamefont {Zhang}},\ }in\ \href@noop
  {} {\emph {\bibinfo {booktitle} {SPIE Astronomical Telescopes+
  Instrumentation}}}\ (\bibinfo {organization} {International Society for
  Optics and Photonics},\ \bibinfo {year} {2014})\ pp.\ \bibinfo {pages}
  {914856--914856}\BibitemShut {NoStop}%
\bibitem [{\citenamefont {Cui}\ and\ \citenamefont
  {Soh}(2011)}]{cui2011analysis}%
  \BibitemOpen
  \bibfield  {author} {\bibinfo {author} {\bibfnamefont {S.}~\bibnamefont
  {Cui}}\ and\ \bibinfo {author} {\bibfnamefont {Y.~C.}\ \bibnamefont {Soh}},\
  }\href@noop {} {\bibfield  {journal} {\bibinfo  {journal} {Optics letters}\
  }\textbf {\bibinfo {volume} {36}},\ \bibinfo {pages} {1692} (\bibinfo {year}
  {2011})}\BibitemShut {NoStop}%
\end{thebibliography}%

\end{document}